\documentclass[aps,prl,letterpaper,superscriptaddress,twocolumn]{revtex4-1}

\usepackage{amstext,amsmath,amssymb,amsthm}
\usepackage{color}

\usepackage{times}

\usepackage[T1]{fontenc}
\usepackage[utf8]{inputenc}
\usepackage{graphicx}
\usepackage[colorlinks]{hyperref}

\newcommand{\beq}{\begin{equation}}
\newcommand{\eeq}{\end{equation}}

\newcommand{\ket} [1] {\vert #1 \rangle}
\newcommand{\bra} [1] {\langle #1 \vert}

\newcommand{\Tr}{\mathop{\mathrm{Tr}}}

\newcommand{\ba}{\begin{align}}
\newcommand{\ea}{\end{align}}
\newcommand{\bea}{\begin{eqnarray}}
\newcommand{\eea}{\end{eqnarray}}

\renewcommand{\O}{O}

\DeclareMathOperator{\var}{var}

\usepackage{tikz}	
\usepackage[hang,small,bf]{caption}  
\usetikzlibrary{backgrounds,fit,decorations.pathreplacing}  
\makeatletter

\@ifundefined{textcolor}{}
{%
 \definecolor{BLACK}{gray}{0}
 \definecolor{WHITE}{gray}{1}
 \definecolor{RED}{rgb}{1,0,0}
 \definecolor{GREEN}{rgb}{0,.4,0}
 \definecolor{BLUE}{rgb}{0,0,1}
 \definecolor{CYAN}{cmyk}{1,0,0,0}
 \definecolor{MAGENTA}{cmyk}{0,1,0,0}
 \definecolor{YELLOW}{cmyk}{0,0,1,0}
 }

\makeatother

\usepackage{bm}
\usepackage{mathtools}

\def\id{I}

\def\1{\mat{\id}}
\def\mat#1{\mathbf{#1}}

\begin{document} 

\title{Quantum algorithms for training Gaussian Processes}
\author{Zhikuan Zhao}
\affiliation{Singapore University of Technology and Design, 8 Somapah Road, Singapore 487372}
\affiliation{Centre for Quantum Technologies, National University of Singapore, 3 Science Drive 2, Singapore 117543}
\author{Jack K. Fitzsimons} 
\affiliation{Department of Engineering Science, University of Oxford, Oxford OX1 3PJ, UK}
\author{Michael A. Osborne}
\affiliation{Department of Engineering Science, University of Oxford, Oxford OX1 3PJ, UK}
\author{Stephen J. Roberts}
\affiliation{Department of Engineering Science, University of Oxford, Oxford OX1 3PJ, UK}
\author{Joseph F. Fitzsimons}
\email{joseph_fitzsimons@sutd.edu.sg}
\affiliation{Singapore University of Technology and Design, 8 Somapah Road, Singapore 487372}
\affiliation{Centre for Quantum Technologies, National University of Singapore, 3 Science Drive 2, Singapore 117543}
\begin{abstract}
Gaussian processes (GPs) are important models in supervised machine learning. Training in Gaussian processes refers to selecting the covariance functions and the associated parameters in order to improve the outcome of predictions, the core of which amounts to evaluating the logarithm of the marginal likelihood ($\mathrm{LML}$) of a given model. $\mathrm{LML}$ gives a concrete measure of the quality of prediction that a GP model is expected to achieve. The classical computation of $\mathrm{LML}$ typically carries a polynomial time overhead with respect to the input size. We propose a quantum algorithm that computes the logarithm of the determinant of a Hermitian matrix, which runs in logarithmic time for sparse matrices. This is applied in conjunction with a variant of the quantum linear system algorithm that allows for logarithmic time computation of the form $\mathbf{y}^TA^{-1}\mathbf{y}$, where $\mathbf{y}$ is a dense vector and $A$ is the covariance matrix. We hence show that quantum computing can be used to estimate the $\mathrm{LML}$ of a GP with exponentially improved efficiency under certain conditions. 
\end{abstract}
\maketitle
\date{today}
\maketitle
The goal of supervised machine learning is to infer a function from a labelled set of input and output example points, known as the training data \cite{mohri2012foundations}. Gaussian processes (GPs) represent an approach to supervised learning that models the underlying functions associated with the outputs in an inference problem as an infinite-dimensional generalisation of a Gaussian distribution. GPs offer a model suitable for capturing a wide range of behaviours governed only by a simple set of parameters \cite{GP}. 

A limiting factor for many applications of machine learning is the computational complexity of the underlying algorithms. This can prove prohibitive when working with large datasets. Machine learning has naturally arisen as a potential application for quantum computation. In recent years, significant progress has been made on quantum algorithms for machine learning tasks from several directions \cite{schuld2015introduction}. Quantum algorithms have been proposed for many of the operations commonly used in machine learning, including matrix inversion \cite{HHL}, principal component analysis \cite{lloyd2014quantum} and clustering \cite{aimeur2013quantum,lloyd2013quantum}. Quantum algorithms for support vector machines \cite{rebentrost2014quantum}, accelerated deep learning \cite{wiebe2014quantum} and pattern classification based on linear regression \cite{schuld2016pattern} have also emerged, and the use of quantum annealers and adiabatic quantum computation has begun to show promise for machine learning tasks \cite{neven2008training,pudenz2013quantum,amin2016quantum}. More recent advances in the field are well summarized in \cite{adcock2015advances}. A quantum algorithm for Gaussian process regression (GPR) has previously been presented in \cite{zhao2015quantum}. The proposed Quantum Gaussian Process Algorithm (QGPA) leverages the exponential speed-up achievable in the quantum linear system algorithm \cite{HHL}, and estimates the mean and variance of the predictive distribution given by the GP.
In this work, we complete the quantum Gaussian process procedure by providing a similarly efficient quantum routine to select the covariance functions and the associated parameters necessary to train a GP model.

An important aspect of many supervised learning approaches amounts to the ability to efficiently select preferred variations of the model, in order to achieve better predictions. A significant amount of attention in machine learning research has rightfully been devoted to this process of model selection which is a somewhat under explored topic in the existing quantum machine learning literature. In the context of GPs, this amounts to choosing a covariance function, known as the GP kernel. In practice, a family of functions are usually considered. The parameters of the family of kernels are referred to as kernel hyper-parameters and a range of optimisers are used in order to tune these hyper-parameters based on the observed data. This process is commonly known as the training of a Gaussian process.

Since model selection typically involves repeated evaluation of certain cost functions that characterise how well a given model is performing on the problem, it generally carries a runtime overhead that scales polynomially with the input size. As QGPA provides a speed-up in computing predictions given a fixed kernel, it is desirable to also have a correspondingly efficient quantum routine for kernel and parameter selection. In particular, it would be desirable to evaluate a measure of the model's performance with a quantum routine that supplements the main learning algorithm. With the above motivation, we propose a quantum approach to improve the efficiency of GP training based on evaluating the logarithm of marginal likelihood ($\mathrm{LML}$) of the Gaussian distribution of the observed data.

We begin by reviewing the basics of Gaussian processes with an emphasis on kernel selection and hyper-parameter tuning. We will then introduce the first algorithm of the paper, namely a quantum log determinant algorithm, which computes the logarithm of the determinant of a Hermitian matrix. The second algorithm builds on the well-known quantum linear system algorithm \cite{HHL} and allows us to compute $\mathbf{y}^TA^{-1}\mathbf{y}$ for sparse matrix $A$ and dense vector $\mathbf{y}$. In conjunction, the two algorithms can be applied to improve the efficiency of evaluating the $\mathrm{LML}$ of a GP model, potentially yielding an exponential improvement in performance. We conclude with a discussion on the potential sources of estimation errors and the practical application of our procedure.

Following the convention of \cite{GP}, we begin by introducing the fundamentals of GPs and how model selection is typically implemented. Consider a supervised learning problem with a set of training data $\mathcal{T}=\{\mathbf{x}_i,y_i\}^{n-1}_{i=0}$, which contains $n$ $d$-dimensional inputs, $\{\mathbf{x}_i\}_{i=0}^{n-1}$, and their corresponding outputs, $\{y_i\}_{i=0}^{n-1}$. We are interested in modelling the underlying function which generated the dataset. We refer to this as the latent function $f(\mathbf{x})$. 
This latent function is related to the output data by 
$$
y = {f}(\mathbf{x}) + \epsilon_{\text{noise}},
$$
 where $\epsilon_{\text{noise}}\sim\mathcal{N}(0,\sigma_n^2)$ represents independent and identically distributed Gaussian noise, with zero mean and variance $\sigma_n^2$. When given a new input, $\mathbf{x}_*$, we aim to have a predictive distribution of ${f}_*={f}(\mathbf{x}_*)$. A GP method models $\{{f}(\mathbf{x_i})\}_{i=0}^{n-1}$ as a joint multi-dimensional Gaussian distribution, which can be completely specified by a mean function, 
 $$m(\mathbf{x})=\mathbb{E}[{f}(\mathbf{x})],$$ 
 and a covariance function (or kernel function), 
 $$k(\mathbf{x},\mathbf{x}^\prime)=\mathbb{E}[({f}(\mathbf{x})-m(\mathbf{x}))({f}(\mathbf{x}^\prime)-m(\mathbf{x}^\prime))],$$ with a fixed set of hyper-parameters. For the sake of simplicity, we can choose to set the GP model to have zero prior mean. An arbitrary parametric prior mean could be equally well learned by the techniques to follow.
We write the predictive distribution of ${f}_*$, conditioned on $\{\mathbf{x}_i,y_i\}^{n-1}_{i=0}$ and the target input $\mathbf{x}_*$, as a multi-variable Gaussian distribution: 
$$
p({f}_*|\mathbf{x}_*,\mathcal{T})\sim\mathcal{N}(\bar{{f}_*},\mathbb{V}[{f}_*]),
$$
where $\bar{{f}_*}$ and $\mathbb{V}[{f}_*]$ denotes the mean and the variance of the Gaussian distribution respectively.

The aim of training is to achieve a better predictive distribution for a given problem by selecting the form of the covariance function and varying its parameters. 
The natural heuristic for the suitability of a supervised learning model is the marginal likelihood, $p(\mathbf{y}|K+\sigma_n^2{I})$, where $K\in\mathbb{R}^{n\times n}$ denotes the covariance matrix between the input points, which is by definition symmetric positive semi-definite. We have employed a vectorised notation for the dataset, such that $\mathbf{y}\in\mathbb{R}^n$ contains entries given by the outputs. Intuitively, we wish to tune the covariance function and respective hyper-parameters by maximising the probability of the observed data given the GP prior.
Since we have chosen the model to have zero prior mean, we can write down the distribution of $\mathbf{y}$ as $\mathbf{y}\sim\mathcal{N}(0,K+\sigma_n^2{I})$. It follows that the logarithm of marginal likelihood $\mathrm{LML}=\log[p(\mathbf{y}|K+\sigma_n^2{I})]$ is then convenient to compute using the following identity \cite{GP}:
\begin{align}
\mathrm{LML} =&-\frac{1}{2}\log\det[K+\sigma^2_n\mathit{I}] -\frac{1}{2}\mathbf{y}^T(K+\sigma_n^2\mathit{I})^{-1}\mathbf{y}\nonumber\\
&-\frac{n}{2}\log 2\pi.\label{eq:lml}
\end{align}
Note that $\sigma_n^2{I}$ is added to account for the fact that although our GP models the latent function as a Gaussian distribution, the output signal is potentially noisy, and hence there is additional variance in the predicted distribution of the output.
The first term of $\mathrm{LML}$ only depends on the covariance matrix with an identity noise entry, and amounts to a penalty on the complexity of the model. It will disfavour models which overfit the dataset. The second term in $\mathrm{LML}$ is the only one which involves the observed output data, and is therefore responsible for indicating how well the model is actually fitting the data. The final term, $-\frac{n}{2}\log 2\pi$ is an easily computable normalisation constant. Therefore only the first two terms in $\mathrm{LML}$ requires quantum algorithms for speed-up.

Classically, the runtime in computing $\mathrm{LML}$ is dominated by computing matrix multiplication and determinant, and hence scales with the data size as $O(n^3)$ in typical implementations. A possible improvement can be achieved with optimised CW-like algorithms \cite{williams2011breaking}, cutting the runtime down to $O(n^{2.373})$, although such scaling is difficult to achieve in practice. Due to this high computational cost, a number of approximate methods have arisen within the machine learning community. Very often GPs are constrained to a have fixed rank in order to make the computation more manageable. In these cases a range of methods can be used to reduce the complexity of training to $O(nr^2)$, where $r$ is the rank of the covariance matrix \cite{quinonero2005unifying}. Unfortunately, this limits the complexity of the functions which can be modelled by the GP and ultimately hinders the model's predictive performance. Other approaches, such as hierarchical matrix factorisation \cite{minden2016fast}, work well in low-dimensional spaces but do not scale well to high dimensional datasets which are often of interest. 

More recently, stochastic trace estimation approaches have become popular \cite{pace2004chebyshev, boutsidis2015randomized}. These methods utilise the equality relationship between the log determinant of a matrix and the trace of the log of the matrix. Using this relationship, the logarithm of a matrix is approximated either by truncating the Taylor series of the matrix logarithm,
or by approximating it using a Chebyshev polynomial approximation of some user specified degree $d$. The advantage of a trace estimation approach is that raising $A$, or $(I - A)$ for that matter, still requires matrix multiplication but the speed-up arises as the product $z^\dagger\log(A)z$ can be computed in $O(n^2)$ for some $z \in \mathbb{R}^{n}$. As such the `probing vectors', $z$, are chosen such that $\mathbb{E}[z^\dagger\log(A)z] = \Tr(\log(A))$ and can be done so in a number of ways \cite{avron2011randomized, MUB}. Note that there are two sources of error which occur in such an approach, namely due to the approximation of $\log(A)$ and due to the stochastic trace estimation. We draw particular attention to these stochastic trace estimation methods as the approach considered here may be seen as an extension of this class of algorithms. Relative to those algorithms, our approach offers both less error, due to an exact representation of $\log(K+\sigma_n^2I)$ to machine precision, and an exponential reduction in computation time over classical algorithms. 

We address the first term in \eqref{eq:lml} by describing a quantum procedure to efficiently sample the eigenvalues of an $n\times n$ Hermitian matrix $A$ uniformly at random, based on phase estimation \cite{kitaev1995quantum,buvzek1999optimal,luis1996optimum,cleve1998quantum}. This can be seen as a finite dimensional analogue of the continuous variable model proposed in \cite{nana}. For simplicity, we will assume that $n=2^N$. The algorithm then proceeds as follows:

\textit{Step 1.} Prepare $N$ qubits in maximally-mixed state, $\frac{1}{n}\sum\limits_{i=1}^{n}\ket{i}\bra{i}$, and store this in a first register. This can be achieved simply by preparing the register in a random computational basis state. Note that a maximally-mixed state is maximally-mixed in any basis, hence we can choose to represent the density matrix for the system it in the eigenbasis $\{\ket{e_i}\}$ of matrix $A$: 
\begin{align}
\frac{1}{n}\sum\limits_{i=1}^{n}\ket{e_i}\bra{e_i}.\nonumber
\end{align}

\textit{Step 2.} Append a second register in a superposition state given by $\frac{1}{\sqrt{T}}\sum\limits_{\tau=1}^{T}\ket{\tau}$, so that the composite system is in the state
\begin{align}
\frac{1}{nT}\sum\limits_{\tau,\tau^\prime=1}^{T}\sum\limits_{i=1}^{n}\ket{e_i}\bra{e_i}\otimes\ket{\tau}\bra{\tau^\prime}.\nonumber
\end{align}
Here $T$ is chosen to be some sufficiently large to ensure accurate phase estimation as described in \cite{buvzek1999optimal}.

\textit{Step 3.} Treating $(-A)$ as a Hamiltonian (which is possible since $A$ is Hermitian), evolution under $(-A)$ for time specified by the second register is simulated on the state stored in the first register. This is achieved by applying the conditional unitary evolution $\sum\limits_{\tau=1}^{T}\mathrm{e}^{iAt_0\tau/T}\otimes\ket{\tau}\bra{\tau}$, where $t_0=O(1/\epsilon)$ is chosen with respect to the $\epsilon$-bounded error required in the algorithm. We thus obtain the state
\begin{align}
\frac{1}{nT}\sum\limits_{\tau,\tau'=1}^{T}\sum\limits_{i=1}^{n}\mathrm{e}^{i\lambda_it_0(\tau-\tau')/T}\ket{e_i}\bra{e_i}\otimes\ket{\tau}\bra{\tau'}. \nonumber
\end{align}

\textit{Step 4.} Perform a quantum Fourier transform of the second register. The resulting estimated eigenvalues of $A$, $\{\lambda_i\}$, are then stored in the second register as a binary bit-string up to a finite precision associated with the phase estimation procedure.  Thus this results in the system being in state
\begin{align}
\frac{1}{n}\sum\limits_{i=1}^{n}\ket{e_i}\bra{e_i}\otimes\ket{\lambda_i}\bra{\lambda_i}.\nonumber
\end{align}

\textit{Step 5.} Measure the second register in computational basis to obtain a random $\lambda_i$. 

This sampling method can then be turned to the task of estimating the log determinant of $A$, by making use of the identity
\begin{align}
\langle \log\lambda_i\rangle = \frac{1}{n}\sum\limits_{i=1}^{n}\log\lambda_i = \frac{1}{n}\Tr[\log(A)] = \frac{1}{n}\log[\det(A)].\nonumber
\end{align}
Hence the desired quantity $\log[\det(A)]$ is given by $n\langle \log\lambda_i\rangle$ which will need to be estimated by sampling eigenvalues of $A$ on repeated runs of the above procedure.  The ``penalty'' term of the $\mathrm{LML}$ can now be estimated using the eigenvalue sampling procedure described here, by setting $A = K+\sigma_n^2\mathit{I}$. 

Next we consider the second term of \eqref{eq:lml}, $\frac{1}{2} \mathbf{y}^TA^{-1}\mathbf{y}$, which is often referred to as the ``data fit'' term and relates the dense outputs $\mathbf{y}$ to the assumed sparse covariance matrix $K$. We will show how a modified version of the quantum linear system algorithm (QLSA) \cite{HHL} can be used in order to calculate this term.
The QLSA operates similarly to phase estimation discussed earlier in this paper, with the addition of an ancilla qubit which is rotated conditioned on the values of $f(\lambda_i)$, the eigenvalues of $A$, in the linear equations $A\ket{\mathbf{x}} = \ket{\mathbf{b}}$, and the non-linear function $f$. In the case of \cite{HHL}, $f$ is simply the inverse of the eigenvalues. Post-selecting this ancilla qubit followed by the reversal of the phase estimation step results in finding $A^{-1}\ket{\mathbf{b}}$ with success probability $\bra{\mathbf{b}}(A^{-1})^\dagger A^{-1}\ket{\mathbf{b}}$. 

As the authors of \cite{HHL} note, the transformation $f(A)$ may be replaced using any computable function $f$. Here we are interested in the case when $f(x) = \frac{1}{\sqrt x}$ rather than the original inversion used. This new function reduces the effect of poor conditioning by a square-root as the success probability of the measurement step is increased as $\sqrt x \geq x$ for all $0 \leq x \leq 1$. We will further make use of a quantum random access memory (QRAM) \cite{RAM} to prepare $\ket{\mathbf{y}}=\frac{\mathbf{y}}{\|\mathbf{y}\|}$ with the state preparation technique described in \cite{zhao2015quantum}, and set $\ket{\mathbf{b}}=\ket{\mathbf{y}}$ as well as $A = K+\sigma_n^2\mathit{I}$. As such the result of the algorithm leads to $A^{-\frac{1}{2}}\ket{\mathbf{y}}$ with success probability $\bra{\mathbf{y}}A^{-1}\ket{\mathbf{y}}$. We apply this modified linear system algorithm in order to compute a Monte Carlo estimate of the data fit term with mean $\mathbf{y}^TA^{-1}\mathbf{y}$ and variance bounded by $\frac{1}{4}\|\mathbf{\mathbf{y}}\|^2\sigma_n^{-2}$.
When $A$ is sparse and well-conditioned, the runtime of sampling from such a distribution is logarithmic in the dimension of $\mathbf{y}$, inherited from the computational cost of QLSA in \cite{HHL}.

The full quantum estimation of $\mathrm{LML}$ is obtained by combining the ``penalty'' and the ``data fit'' terms. For the purpose of GP training, we are concerned with estimating the variation, $\delta\mathrm{LML}$, with respect to a training step, where the prefix $\delta$ denotes the change in a quantity between steps.
The figure of merit for the estimation error is the relative variance, as it quantifies the amount of dispersion between the estimated and the actual variation of $\mathrm{LML}$. In order to demonstrate the quantum advantage, it is therefore necessary to show that the relative variance with respect to a change in hyper-parameter, $\delta\theta$, does not scale up with $n$. We consider the following,
\begin{align}
\frac{\var\left[\delta \mathrm{LML}\right]}{\left[\delta \mathrm{LML}\right]^2}=&\frac{ \var\left[\log[\det(A)]\right]+\var\left[ \mathbf{y}^TA^{-1}\mathbf{y} \right] }{\left[\frac{\partial}{\partial\theta}\left(\log[\det(A)]+\mathbf{y}^TA^{-1}\mathbf{y}\right)\delta\theta\right]^2}\nonumber\\
\le&\frac{n^2 \left(\var\left[\log \lambda_i\right] + \frac{1}{4}\left<y_i^2\right>\sigma_n^{-2}\right)}{\left[\frac{\partial}{\partial\theta}\left(\sum_i\log\lambda_i+\sum_i|\gamma_i|^2\lambda_i^{-1}\right)\delta\theta\right]^2}\nonumber\\
\le&\frac{\left<(\log\lambda_i)^2\right>+\frac{1}{4}\left<y_i^2\right>\sigma_n^{-2}}{\left<\delta\lambda_i/\lambda_i+\delta\left(|\gamma_i|^2/\lambda_i \right)\right>^2}, \label{revar}\nonumber
\end{align}
where we have written the $\mathbf{y}$ as a linear combination of the eigenvectors, $\mathbf{e}_i$ of $A$, such that $\mathbf{y}=\sum_i\gamma_i\mathbf{e}_i$, and $\mathbf{y}^TA^{-1}\mathbf{y}=\sum_i|\gamma_i|^2\lambda_i^{-1}$. The expectation value notation is used to denote the average over all choices of $i$. Hence the relative variance in estimating the variation of $\mathrm{LML}$ with respect to a training step has no explicit dependence on $n$.

An important component in our algorithm is the procedure of Hamiltonian simulation which amounts to exponentiating $A$ in order to construct a unitary operator $\mathrm{e}^{-iAt}$. A general technique to achieve this is based on a Hamiltonian simulation method described in \cite{berry2015hamiltonian} which is based on Szegedy quantum walks \cite{berry2009black}. This method enables the exponentiation of an $n\times n$ Hermitian matrix in $\tilde{O}(s\log n)$ given oracle access to the matrix elements, where $s$ denotes the sparseness of $A$, such that there are at most $s$ non-zero elements in each column/row. Here we have used the notational shorthand convention $\tilde{O}(x)$ to denote $O(x\log^k(x))$ for any constant $k$, such that slower growing contributions are omitted. When dealing with non-sparse but low-rank matrices, another technique of Hamiltonian simulation involving density matrix exponentiation \cite{lloyd2014quantum} can potentially be applied. Note that the covariance matrices are by definition symmetric, real and positive semi-definite, and therefore have very similar mathematical structure to the density matrix representation of quantum states. Hence this seminal technique of density matrix exponentiation potentially allows us to implement $\mathrm{e}^{-iAt}$ in $\tilde{O}(\log n)$ time, even if the matrix is not sparse. However it should be noted that the covariance matrix needs to be normalised to have unit trace for the application of density matrix exponentiation. This pre-processing can be done efficiently if one can exploit the analytical structure of the covariance matrix. It should also be noted that if the eigenvalues of the covariance matrix are relatively uniform, the time required to implement the unitary for a complete cycle will scale as $O(n)$. Hence applying density matrix exponentiation is most effective when the covariance matrix is approximately low-rank \cite{lloyd2014quantum}. To keep our analysis fully general, we will include the linear sparseness dependence in our counting of runtime.

The optimised phase estimation procedure \cite{buvzek1999optimal,luis1996optimum} comes with an error, $\epsilon_{\lambda_i}$, which scales as $O(1/t_0)$ in estimating each $\lambda_i$. This implies the error associated with the logarithm of a single eigenvalue scales as
$\epsilon=\left|\frac{d\log\lambda_i}{d\lambda_i}\epsilon_{\lambda_i}\right|=\O\left(\frac{1}{\lambda_it_0}\right)$.
Furthermore, in the context of GP training, there generally exists a $\sigma_n^2 \mathit{I}$ noise contribution to the covariance matrix, due to uncertainty in the observed data. Thus, in general, we have the minimum eigenvalue, $\lambda_{\text{min}}\ge\sigma_n^2$. Hence, the total bounded-error single-run of the algorithm takes time scaling logarithmically in $n$ as $t=\tilde{\O}\left(\frac{s\log n}{\sigma_n^2\epsilon}\right)$.

Due to the linear sparseness dependence from the Hamiltonian simulation step, our algorithm performs best when the covariance matrix is some constant $s$-sparse, in which case our algorithm provides an exponential speed-up over the classical GP training procedure. Such sparsely constructed GPs have found applications in a range of interesting problems, especially when large size datasets are involved \cite{sparse2009}. For examples, a sparse Gaussian process is used to construct a unified framework for robotic mapping in \cite{robotic}. It was also applied to realistic action recognition problems in \cite{recognition}.
When applied to a non-sparse covariance matrix, the density matrix exponentiation procedure \cite{lloyd2014quantum} can still lead to a logarithmic time algorithm if the matrix has a low-rank structure. In other cases, a singular value estimation scheme which circumvents the Hamiltonian simulation step can be applied to achieve a runtime that scales as $\tilde{\O}(\sqrt{n}\log n)$, assuming the spectral norm of $A$ is bounded by a constant \cite{dense,Kerenidis2016,GD}. This provides a polynomial speed-up over the best classical counterpart. 

Returning to the comparison with classical stochastic trace estimation methods, it is clear that the quantum algorithm offers a precise method to compute $\log(A)$ rather than either the truncated Taylor series or Chebyshev polynomial approximations. When measurements of the second register are taken, a single $\log(\lambda_i)$ is computed and hence our proposed approach can be seen as quantum stochastic trace estimation. The main advantage, however, comes from the reduction in computation time from polynomial to sub-linear. A natural question which arises is whether the complete GP training can scale sub-linearly in $n$, since if not, an exponential improvement in computing the $\mathrm{LML}$ in each step would yield only a polynomial improvement in precision. Note that the number of hyper-parameters is dependent only on the kernel, and thus independent of the number of data points. Provided we are working to constant precision, the number of optimisation steps which require $\mathrm{LML}$ computation is upper bounded by a constant. 

We have shown a quantum algorithm that improves the efficiency of calculating $\mathrm{LML}$ from a classical $\O(n^3)$ to a logarithmic scaling with respect to the size of input under certain conditions. If either the structure of the covariance matrix is constant $s$-sparse or approximately low-rank (such that the density matrix exponentiation scheme can be efficiently applied), our algorithm provides an exponential speed-up. Even in cases when the Hamiltonian simulation step necessarily consumes a $\tilde{O}(n\log n)$ time overhead, this quantum algorithm still achieves a polynomial speed-up over the best known classical approach to train full-rank GPs.

\textit{Acknowledgements}--- The authors thank Mahboobeh Houshmand Kaffashian, Nana Liu and Liming Zhao for useful comments on the manuscript. JFF acknowledges support from the Ministry of Education Singapore, and the Air Force Office of Scientific Research under AOARD grant no. FA2386-15-1-4082. This material is based on research supported by the Singapore National Research Foundation under NRF Award No. NRF-NRFF2013-01.
 
\bibliographystyle{apsrev}
\bibliography{training}
\end{document}